\documentclass[conference]{IEEEtran}
\usepackage{xcolor,soul,framed} 

\colorlet{shadecolor}{yellow}
\usepackage[pdftex]{graphicx}
\graphicspath{{../pdf/}{../jpeg/}}
\DeclareGraphicsExtensions{.pdf,.jpeg,.png}

\usepackage[cmex10]{amsmath}
\usepackage{cite}
\usepackage{rotating}
\usepackage{hyperref} 
\usepackage{tablefootnote}
\usepackage{array}
\usepackage{mdwmath}
\usepackage{amsmath}
\usepackage{amsfonts}
\usepackage{amssymb}
\usepackage{mdwtab}
\usepackage{eqparbox}
\usepackage{url}
\usepackage{xcolor}
\usepackage{comment}
\usepackage{footnote}
\usepackage{multirow}
\usepackage{caption}
\usepackage{algorithm, algorithmicx}
\usepackage{algpseudocode}

\usepackage{float}
\usepackage{placeins}

\usepackage{url} 
\usepackage{array, makecell}
\usepackage{booktabs}
\usepackage{tikz}

\usepackage{subcaption}

\usepackage{algpseudocode}
\ifCLASSINFOpdf
  
\else
 
\fi

\begin{document}
\bstctlcite{IEEEexample:BSTcontrol}
%
\title{Toward a Unified Semantic Loss Model for Deep JSCC-based Transmission of EO Imagery   \\
}

    \author{\IEEEauthorblockN{Ti Ti Nguyen, Thanh-Dung Le, Vu Nguyen Ha, Duc-Dung Tran, 
    Hung Nguyen-Kha, Dinh-Hieu Tran, \\
    Carlos L. Marcos-Rojas, Juan C. Merlano-Duncan, and Symeon Chatzinotas 
    }\\
    \vspace{-2mm}
    \IEEEauthorblockA{\textit{Interdisciplinary Centre for Security, Reliability and Trust (SnT), University of Luxembourg, Luxembourg} 
	}
    }

\maketitle

\begin{abstract}

Modern Earth Observation (EO) systems increasingly rely on high-resolution imagery to support critical applications such as environmental monitoring, disaster response, and land-use analysis. Although these applications benefit from detailed visual data, the resulting data volumes impose significant challenges on satellite communication systems constrained by limited bandwidth, power, and dynamic link conditions. 
To address these limitations, this paper investigates Deep Joint Source–Channel Coding (DJSCC) as an effective SC paradigm for the transmission of EO imagery. We focus on two complementary aspects of semantic loss in DJSCC-based systems. First, a reconstruction-centric framework is evaluated by analyzing the semantic degradation of reconstructed images under varying compression ratios and channel signal-to-noise ratios (SNR). Second, a task-oriented framework is developed by integrating DJSCC with lightweight, application-specific models (e.g., EfficientViT), with performance measured using downstream task accuracy rather than pixel-level fidelity. 
Based on extensive empirical analysis, we propose a unified semantic loss framework that captures both reconstruction-centric and task-oriented performance within a single model. This framework characterizes the implicit relationship between JSCC compression, channel SNR, and semantic quality, offering actionable insights for the design of robust and efficient EO imagery transmission under resource-constrained satellite links.

\end{abstract}


\IEEEpeerreviewmaketitle

\section{Introduction}
\label{sec:Intro}

Modern Earth Observation (EO) systems increasingly rely on high-resolution imagery to support a wide range of applications, including environmental monitoring, disaster response, agriculture, and land-use analysis \cite{leyva2023satellite}.  Recent advances in machine learning (ML) have enhanced EO by enabling efficient analysis of complex datasets, improving accuracy and predictive insights \cite{fontanesi2023artificial,affek2024survey}.
However, the growing resolution and acquisition frequency of EO sensors have led to an unprecedented increase in data volume, posing significant challenges for satellite communication systems. In practical EO missions, especially those involving Low Earth Orbit (LEO) satellites, the transmission of high-resolution imagery is constrained by limited bandwidth, strict power budgets, and dynamically varying channel conditions. As a result, raw EO data often cannot be transmitted reliably or in a timely manner to ground stations equipped with high-capacity AI processors, limiting the effectiveness of downstream EO applications \cite{chou2023edge}.

To address these challenges, semantic communication (SC) has emerged as a transformative paradigm, shifting traditional bit-oriented transmission toward conveying data meaning, thereby reducing overhead and latency in bandwidth-constrained environments \cite{yang2022semantic,chou2024air}. Among SC approaches, deep joint source–channel coding (DJSCC) has shown strong potential by enabling end-to-end learned mappings that jointly optimize compression and transmission robustness over noisy channels \cite{bourtsoulatze2019deep}.
However, deploying DJSCC in EO systems introduces new challenges. Unlike conventional communication systems, whose performance can be optimized using well-defined metrics such as bit error rate or signal-to-noise ratio, SC performance depends on how transmission impairments affect application-level outcomes. Recent works \cite{huang2025d, wang2025joint, ding2023joint} investigate semantic loss models; however, the implicit and highly coupled relationships among DJSCC compression parameters, channel conditions, and semantic quality remain poorly understood. Specifically, \cite{huang2025d} models distortion in SC systems as the sum of two independent components, namely source distortion and channel distortion. The work in \cite{wang2025joint} adopts a generalized logistic function to approximate semantic similarity; however, this formulation cannot fully capture the two-dimensional characteristics of SC systems. In contrast, \cite{ding2023joint} adopts a relatively simple sum of exponential term to approximate the average data reconstruction error. Overall, the lack of tractable and expressive models complicates system design, particularly when balancing image reconstruction quality against downstream task performance.

In this work, we investigate two complementary aspects of deploying DJSCC for EO systems. First, a reconstruction-centric DJSCC framework is analyzed, where performance is measured in terms of the visual fidelity of reconstructed images under varying compression ratios and channel signal-to-noise ratios (SNRs). Second, a task-oriented DJSCC framework is developed by integrating DJSCC with lightweight, task-specific machine learning models, such as EfficientViT, deployed at computationally capable ground stations. In this setup, performance is evaluated using application-level metrics that reflect the utility of the received data for EO tasks, rather than pixel-level similarity.

Building on our prior work on SC using the DVB-S2 standard \cite{nguyen2025semantic}, we extend the analysis to systems employing DJSCC. Leveraging extensive empirical evaluations on real-world EO data, we propose a unified data-driven semantic loss model that captures both reconstruction-centric and task-oriented performance within a common framework. Specifically, we utilize the EuroSAT dataset \cite{helber2019eurosat} for land use and land cover classification, and evaluate DJSCC performance with and without EfficientViT \cite{liu2023EfficientViT} under varying source compression and channel conditions. The proposed model empirically characterizes the relationship among DJSCC compression settings, channel SNR, and semantic effectiveness. This unified semantic performance framework provides practical guidance for designing robust, bandwidth-efficient, and task-aware EO communication systems.

\begin{figure}[t] 
    \centering
    \includegraphics[width=0.46\textwidth]{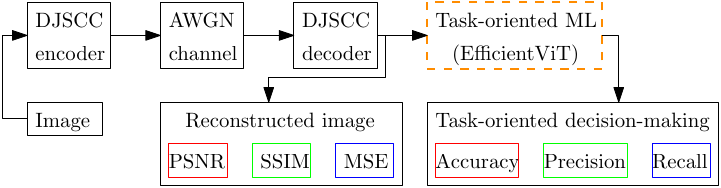}
    \captionsetup{font=small}
    \caption{Block diagram of SC with DJSCC.}
    \label{fig:sys_block_diagram}
    \vspace{- 0.5 cm}
\end{figure}
\section{Data Acquisition for Semantic-Loss Modeling}
\subsection{DJSCC-based Transmission for EO Systems}
\label{sec_2b}

In this paper, we explore the performance of employing DJSCC in transmitting the EO imagery from satellites to the ground.
Specifically, images captured by EO satellites (i.e., the original dataset) are transmitted to ground stations for processing on co-located servers. The images are assumed to be transmitted using the DJSCC technique \cite{bourtsoulatze2019deep}. 
The images are then reconstructed at the ground station.
Classification is then exploited as a well-known application using these reconstructed images for the performance evaluation. 

Unlike traditional communication systems that rely on separate source coding (e.g., JPEG2000) and channel coding (e.g., LDPC or Turbo codes), DJSCC leverages deep neural networks to jointly learn both encoding and decoding functionalities in a single end-to-end model. This enables DJSCC to more effectively preserve semantic information under bandwidth and channel constraints, particularly in non-ideal communication environments where the separation theorem does not hold.

As illustrated in Fig.~\ref{fig:sys_block_diagram}, performance degradation in SC systems using DJSCC stems from two primary sources: (1) source distortion introduced by the encoder–decoder compression process, and (2) channel distortion caused by noisy wireless transmission. To analyze and quantify these effects, we introduce two key parameters: (a) the compression ratio $\rho$, defined as the ratio of the size of the original image to the size of the encoded feature vector, and (b) the signal-to-noise ratio (SNR) $\gamma$, which characterizes the physical channel quality.

\begin{table}[!t]
    \centering
    \caption{\small\textsc{Empirical Results Using DJSCC.}}
    \label{tab:results_DJSCC}
    \begin{tabular}{@{}r*{5}{>{\centering\arraybackslash}p{1.1cm}}@{}}
        \toprule
        \( \gamma \backslash \rho \) & 2 & 4 & 6 & 8 & 12 \\ 
        \midrule
        \multicolumn{6}{c}{PSNR}\\
        \midrule
        8  dB & 35.1659 & 34.5505 & 34.0580 & 33.6110 & 33.2048 \\ 
        4  dB & 34.6355 & 33.7960 & 33.1808 & 32.7066 & 32.1396 \\ 
        0  dB & 33.9541 & 32.9471 & 32.2797 & 31.6707 & 30.9957 \\ 
        -1 dB & 33.6718 & 32.6668 & 31.9636 & 31.4396 & 30.7259 \\ 
        -2 dB & 33.3420 & 32.4279 & 31.7131 & 31.0607 & 30.3161 \\ 
        -3 dB & 33.0512 & 32.0707 & 31.3065 & 30.7091 & 29.9191 \\ 
        -4 dB & 32.7076 & 31.7420 & 30.9968 & 30.3004 & 29.4664 \\ 
        -5 dB & 32.4621 & 31.3836 & 30.5894 & 29.9658 & 29.0585 \\ 
        -6 dB & 32.0890 & 31.0723 & 30.2680 & 29.5925 & 28.6726 \\ 
        -7 dB & 31.8650 & 30.7567 & 29.9635 & 29.2635 & 28.3096 \\ 
        \midrule
        \multicolumn{6}{c}{SSIM}\\
        \midrule
        8 dB  & 93.9029 & 93.0065 & 92.0605 & 90.8419 & 89.3904 \\ 
        4 dB  & 92.8318 & 91.0260 & 89.9122 & 88.5497 & 86.9133 \\ 
        0 dB  & 91.4196 & 88.2488 & 86.1295 & 84.2376 & 81.8966 \\ 
        -1 dB & 90.7393 & 87.6952 & 85.3222 & 83.4208 & 80.6235 \\ 
        -2 dB & 90.2920 & 86.7390 & 84.0706 & 81.7849 & 78.9081 \\ 
        -3  dB& 89.2889 & 85.2669 & 82.4933 & 80.2096 & 76.8448 \\ 
        -4 dB & 88.2929 & 83.8135 & 80.8698 & 78.0683 & 74.5515 \\ 
        -5 dB & 86.9725 & 82.3352 & 78.8621 & 76.3074 & 72.1772 \\ 
        -6 dB & 85.5012 & 80.6746 & 77.0950 & 74.1351 & 69.7947 \\ 
        -7 dB & 84.3367 & 78.9294 & 75.2783 & 72.1513 & 67.5156 \\ 
        \midrule
        \multicolumn{6}{c}{MSE}\\
        \midrule
        8 dB  & 30.1747 & 35.2691 & 40.1839 & 47.5344 & 58.4974 \\ 
        4 dB & 36.4795 & 48.8558 & 55.1101 & 63.4826 & 76.4652 \\ 
        0  dB& 44.4293 & 64.2171 & 77.8224 & 88.3959 & 106.3604 \\ 
        -1 dB & 49.6283 & 65.6896 & 78.7179 & 90.9191 & 111.2500 \\ 
        -2 dB & 50.5202 & 71.9030 & 86.6523 & 102.2033 & 123.2206 \\ 
        -3 dB & 57.5215 & 80.8606 & 97.6941 & 111.2566 & 134.7072 \\ 
        -4 dB & 63.0393 & 89.4421 & 106.4504 & 126.8159 & 151.5233 \\ 
        -5 dB  & 70.4309 & 99.1093 & 122.2410 & 136.6730 & 167.8200 \\ 
        -6 dB & 80.5933 & 110.0527 & 134.3242 & 153.9908 & 185.4238 \\ 
        -7 dB  & 87.7364 & 123.5263 & 147.7795 & 170.0484 & 204.4302 \\ 
        \bottomrule
    \end{tabular}
\end{table}

\begin{table}[!t]
    \centering
    \caption{\small\textsc{Empirical  Task-level results (\%) using  EfficientViT-DJSCC.}}
    \label{tab:results_EfficientViT_DJSCC}
    \begin{tabular}{@{}r*{5}{>{\centering\arraybackslash}p{1.1cm}}@{}}
        \toprule
        \( \gamma \backslash \rho \) & 2 & 4 & 6 & 8 & 12 \\ 
        \midrule
        \multicolumn{6}{c}{Accuracy}\\
        \midrule
        8 dB   & 97.7051 & 97.2967 & 96.9067 & 96.5121 & 95.8632 \\ 
4 dB   & 97.2928 & 96.5092 & 96.0357 & 95.3670 & 94.3556 \\ 
0 dB   & 97.0890 & 95.6177 & 94.5010 & 93.4146 & 91.3953 \\ 
-1 dB  & 96.8802 & 95.3873 & 93.9642 & 92.7888 & 90.1026 \\ 
-2 dB  & 96.6108 & 94.8307 & 93.1518 & 91.5226 & 88.5643 \\ 
-3 dB  & 96.0471 & 93.7941 & 92.0042 & 90.1831 & 86.5272 \\ 
-4 dB  & 95.4615 & 92.9421 & 90.5939 & 88.3597 & 84.3994 \\ 
-5 dB  & 94.7738 & 91.9048 & 89.0595 & 86.6307 & 81.8966 \\ 
-6 dB  & 93.9276 & 90.6833 & 87.5220 & 84.6269 & 79.1926 \\
        \midrule
        \multicolumn{6}{c}{Precision}\\
        \midrule
        8 dB   & 97.6660 & 97.3024 & 96.9487 & 96.5824 & 95.9765 \\ 
4 dB   & 97.2554 & 96.5211 & 96.0831 & 95.4476 & 94.4837 \\ 
0 dB   & 96.8145 & 95.6488 & 94.5692 & 93.5293 & 91.5745 \\ 
-1 dB  & 96.8520 & 95.3238 & 94.0368 & 92.8982 & 90.2869 \\ 
-2 dB  & 96.5923 & 94.8722 & 93.2397 & 91.6484 & 88.7577 \\ 
-3 dB  & 96.0325 & 93.8358 & 92.1079 & 90.3163 & 86.7739 \\ 
-4 dB  & 95.4552 & 92.9980 & 90.7076 & 88.5351 & 84.6581 \\ 
-5 dB  & 94.7731 & 91.9842 & 89.1803 & 86.8053 & 82.1347 \\ 
-6 dB  & 93.9442 & 90.7627 & 87.6891 & 84.8078 & 79.4080 \\  
        \midrule
        \multicolumn{6}{c}{Recall}\\
        \midrule
        8 dB   & 97.6127 & 97.2508 & 96.8902 & 96.5088 & 95.8397 \\ 
4 dB   & 97.2018 & 96.4655 & 96.0222 & 95.3674 & 94.3372 \\ 
0 dB   & 96.7542 & 95.5818 & 94.4970 & 93.4260 & 91.3905 \\ 
-1 dB  & 96.7982 & 95.1904 & 93.9644 & 92.8050 & 90.1037 \\ 
-2 dB  & 96.5321 & 94.8024 & 93.1568 & 91.5443 & 88.5721 \\ 
-3 dB  & 95.9718 & 93.7701 & 92.0142 & 90.2106 & 86.5420 \\ 
-4 dB  & 95.3896 & 92.9224 & 90.6089 & 88.3929 & 84.4213 \\ 
-5 dB  & 94.7052 & 91.8891 & 89.0794 & 86.6693 & 81.9252 \\ 
-6 dB  & 93.8619 & 90.6713 & 87.5462 & 84.6706 & 79.2274 \\ 
        \bottomrule
    \end{tabular}
\end{table}

\subsection{Original Dataset and DJSCC Implementation}
The public benchmark for land use and land cover classification dataset, EuroSAT \cite{helber2019eurosat}, will be used in this study. This is a large-scale benchmark dataset designed explicitly for land use and land cover classification derived from Sentinel-2 satellite imagery. It comprises 27,000 geo-referenced labeled images, each measuring 64x64 pixels and spanning 13 spectral bands. The dataset is categorized into 10 classes. Each
class contains 2000–3000 images, including industrial buildings, residential buildings, annual crops, permanent crops, rivers, seas and lakes, Herbaceous vegetation, highways, pastures, and forests across Europe. Due to its compact image size and diverse class representation, EuroSAT is particularly well-suited for developing and evaluating DL models intended for onboard satellite processing in EO missions. This makes it a valuable resource for applications such as real-time environmental monitoring, disaster response, and precision agriculture. 

To evaluate SC performance, we implement the DJSCC framework proposed in \cite{bourtsoulatze2019deep} under different compression ratios $\rho$ and SNR $\gamma$ are considered. Specifically, we consider $\rho \in \{ 2, 4, 6, 8, 12\}$ and $\gamma \in \{ -6, -5, -4, -3, -2, -1, 0, 4, 8\}$ (dB). Furthermore, motivated by recent findings in \cite{le2024board}, which identify EfficientViT \cite{liu2023EfficientViT} as a highly effective Vision Transformer for EO image classification due to its strong trade-off between performance and computational efficiency, we adopt EfficientViT as the downstream classifier in our evaluation. 
Accordingly, for each combination of compression ratio $\rho$ and SNR $\gamma$, we apply the DJSCC framework with and without EfficientViT. Our empirical results (shown in Tables~\ref{tab:results_DJSCC} and ~\ref{tab:results_EfficientViT_DJSCC}) confirm that both $\rho$ and $\gamma$ jointly influence semantic performance metrics, including both reconstruction fidelity and task-oriented accuracy. Note that \textit{peak signal-to-noise ratio (PSNR), mean squared error (MSE), and structural similarity index measure (SSIM)} are used to assess image reconstruction quality, whereas \textit{accuracy, precision, and recall} are considered for evaluating task performance.

\begin{algorithm}[!t]
\caption{\small\textsc{Gradient Descent for SC Modeling}}\label{alg1}
\small
\begin{algorithmic}[1]
\State \textbf{Initialize:} Set parameters $\boldsymbol{\mu}_{i}$ and learning rates $\alpha_i, i=0:6$, number of terms $N_{\sf c}$,  number of iterations $N$, measured data matrix $\boldsymbol{Y}$ and set of control parameters $\boldsymbol{\rho}$ and $\boldsymbol{\gamma}$.
\State Initialize gradients: ${\sf{Grad}}_{\mu_0} \gets 0$, and ${\sf{Grad}}_ {\mu_i} \gets \boldsymbol{0}$, for $i=1:6$.
\For{$\text{iter} = 1$ to $N$}        
    \State Compute  $\xi$ as defined in \eqref{eq_2dfit}.
    \State Compute error matrix: $\Psi \gets \boldsymbol{Y} - \xi$.
    \For{$i = 1$ to $\text{length}(s)$}
        \For{$j = 1$ to $\text{length}(q)$}
            \For{$k = 1$ to $N_{\sf c}$}
                \State Compute $\sigma_k \gets \frac{1}{1 + \exp(-\mu_{3,k} \gamma_i - \mu_{4,k})}$.
                \State Compute $\eta_k \gets \exp(\mu_{5,k} \rho_j) $.
                \State Compute $\beta_k \gets \eta_k + \mu_{6,k} \rho_j$.
                \State {Update gradients: (where $\epsilon_{ij} = [\Psi]_{i,j}$)\\
                \hspace{2 cm} ${\sf{Grad}}_{\mu_{1,k}} \gets {\sf{Grad}}_{\mu_{1,k}} - 2 \epsilon_{ij} \beta_k$. \\
                \hspace{2 cm} ${\sf{Grad}}_{\mu_{2,k}} \gets {\sf{Grad}}_{\mu_{2,k}} - 2  \epsilon_{ij} \beta_k \sigma_k$.\\
                \hspace{2 cm} ${\sf{Grad}}_{\mu_{3,k}} \gets {\sf{Grad}}_{\mu_{3,k}} $\\
                \hspace{4 cm} $ - 2  \epsilon_{ij}  \beta_k \mu_{2,k} \sigma_k (1 - \sigma_k) \gamma_i$. \\
                \hspace{2 cm} ${\sf{Grad}}_{\mu_{4,k}} \gets {\sf{Grad}}_{\mu_{4,k}}  $\\  
                \hspace{4 cm}$ - 2 \epsilon_{ij} \beta_k \mu_{2,k} \sigma_k (1 - \sigma_k).$\\
                \hspace{2 cm} ${\sf{Grad}}_{\mu_{5,k}} \gets {\sf{Grad}}_{\mu_{5,k}} $\\  
                \hspace{4 cm}$ - 2  \epsilon_{ij}  \eta_k \rho_j (\mu_{1,k} + \mu_{2,k}  \sigma_k).$}\\
                \hspace{2 cm} ${\sf{Grad}}_{\mu_{6,k}} \gets {\sf{Grad}}_{\mu_{6,k}} $\\  
                \hspace{4 cm}$ - 2  \epsilon_{ij}   \rho_j (\mu_{1,k} + \mu_{2,k}  \sigma_k).$
            \EndFor
            \State ${\sf{Grad}}_{\mu_0} \gets {\sf{Grad}}_{\mu_0} - 2 \epsilon_{ij}.$
        \EndFor
    \EndFor
    \State Update: $\mu_{0} \gets \mu_{0} - \alpha_0  {\sf{Grad}}_{\mu_0}$
    \State Update: $\mu_{i,k} \gets \mu_{i,k} - \alpha_i  {\sf{Grad}}_{\mu_{i,k}}$, $i=1:6$, $k=1:N_{\sf c}$.    
\EndFor
\State \textbf{Output:} Final parameters $\boldsymbol{\mu}_i$, $i=0:6$.
\end{algorithmic}
\normalsize
\end{algorithm}

\begin{figure*}[t]
\begin{equation}
\xi = \mu_{0}
+ \sum_{i=1}^{N_{\sf c}}
\left(
\mu_{1,i}
+ \frac{\mu_{2,i}}{1 + \exp\!\left(-\mu_{3,i} \gamma - \mu_{4,i}\right)}
\right)
\left(\exp(\mu_{5,i} \rho) + \mu_{6,i} \rho\right).
\label{eq_2dfit}
\end{equation}
\end{figure*}

\section{Methodology - Curve Fitting-based Modeling}


To the best of our knowledge, there is no theoretical result in the literature for evaluating the EO application objective, particularly in the context of losses arising from source information reduction and wireless transmission. 
In the absence of theoretical models, we employ a practical data-fitting approach to represent the EO application objective as a non-linear function of $\rho$ and $\gamma$. Generally, choosing data-fitting models to optimize system parameters can impact system performance \cite{nguyen2019joint}. A more accurate fitting model typically leads to a more reliable and effective design.


\begin{figure*}[t]
    \centering
    \begin{subfigure}[t]{0.3\textwidth}
        \centering
        \includegraphics[width=\linewidth]{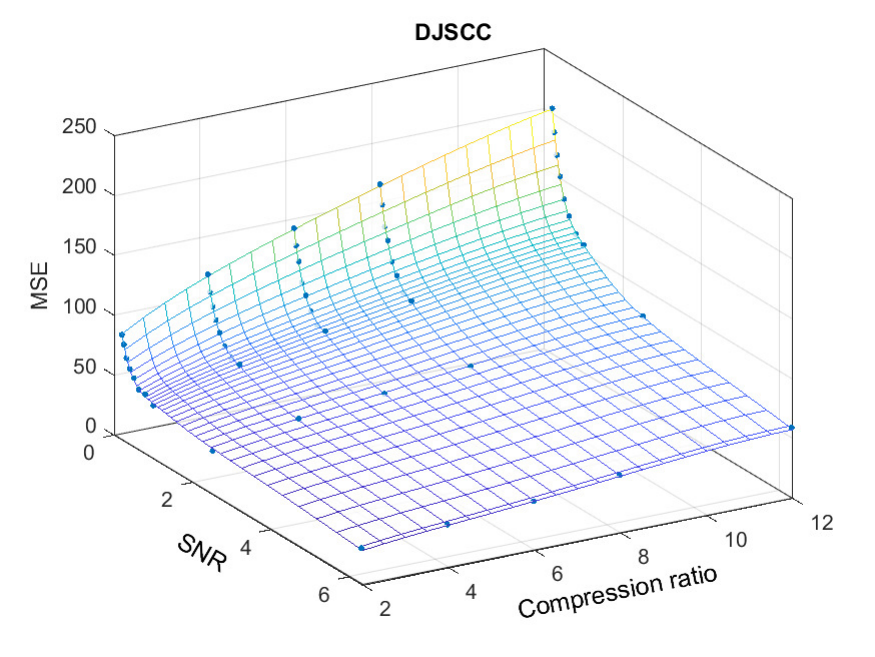}
        \caption{}
        \label{fig:sub1}
    \end{subfigure}
    \hfill
    \begin{subfigure}[t]{0.3\textwidth}
        \centering
        \includegraphics[width=\linewidth]{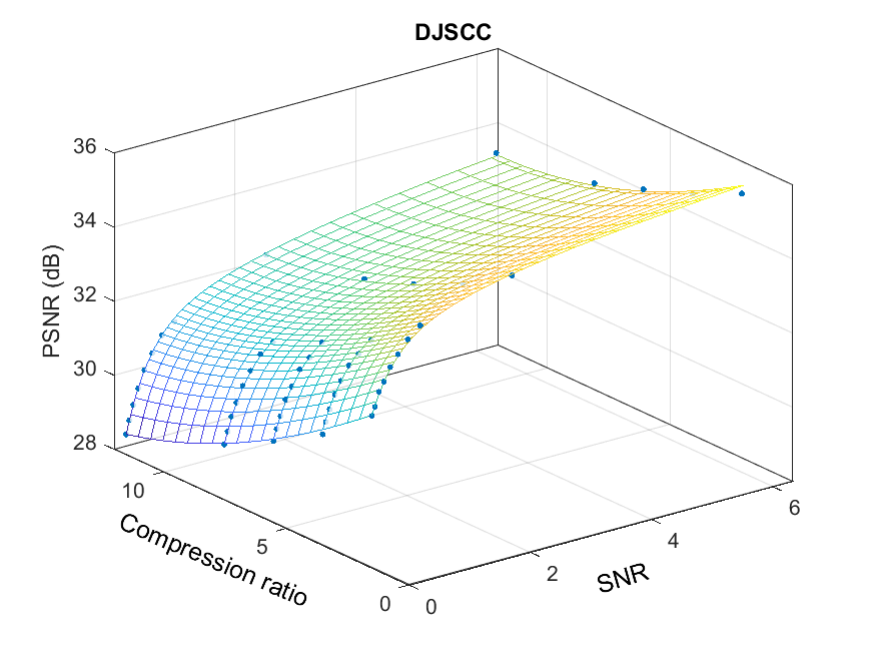}
        \caption{}
        \label{fig:sub2}
    \end{subfigure}
    \hfill
    \begin{subfigure}[t]{0.3\textwidth}
        \centering
        \includegraphics[width=\linewidth]{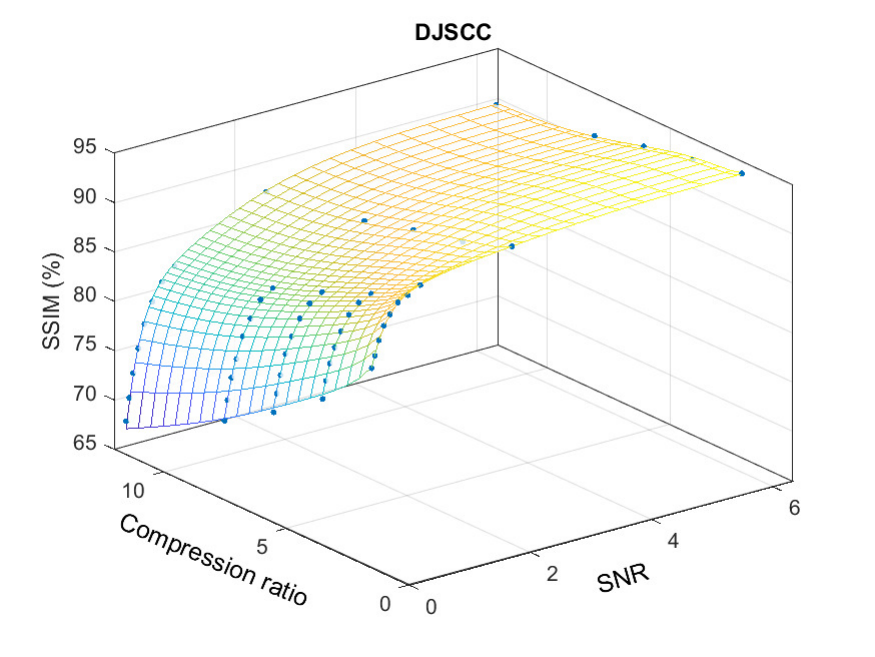}
        \caption{}
        \label{fig:sub3}
    \end{subfigure}

    \caption{Curve fitting models for the DJSCC-AWGN case.}
    \label{fig:three_subfigs}
\end{figure*}

\begin{figure*}[t]
    \centering
    \begin{subfigure}[t]{0.3\textwidth}
        \centering
        \includegraphics[width=\linewidth]{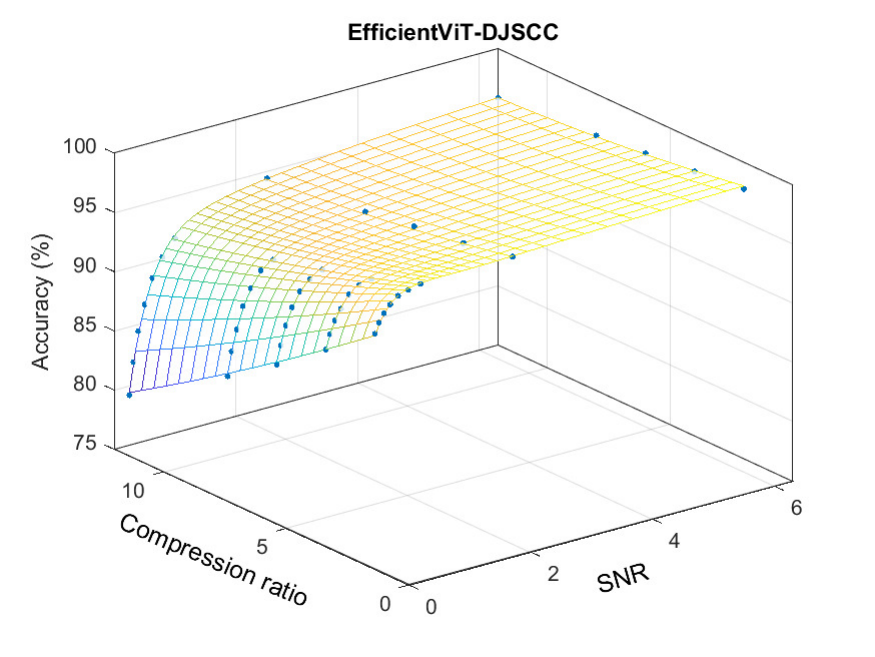}
        \caption{}
        \label{fig:sub1}
    \end{subfigure}
    \hfill
    \begin{subfigure}[t]{0.3\textwidth}
        \centering
        \includegraphics[width=\linewidth]{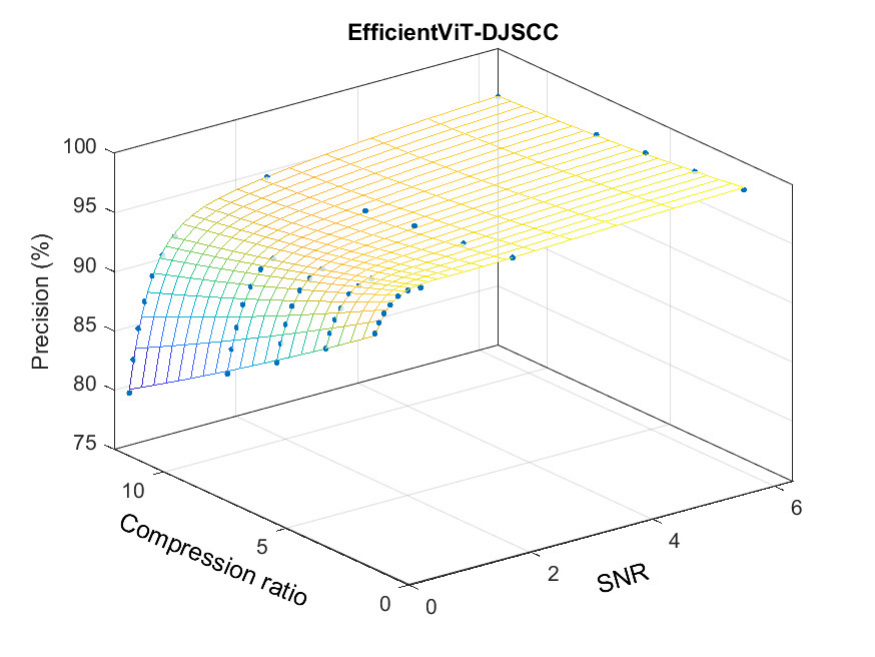}
        \caption{}
        \label{fig:sub2}
    \end{subfigure}
    \hfill
    \begin{subfigure}[t]{0.3\textwidth}
        \centering
        \includegraphics[width=\linewidth]{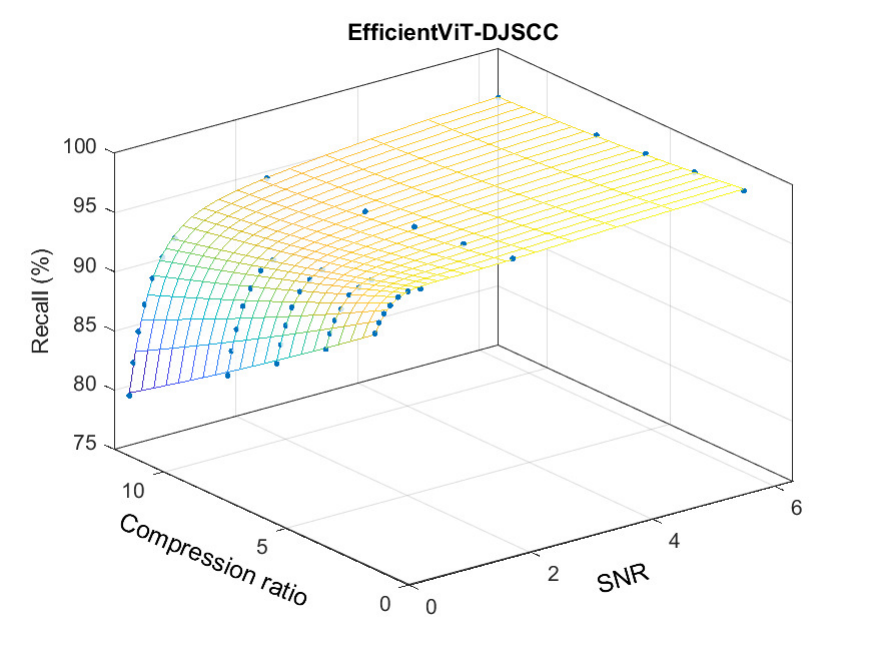}
        \caption{}
        \label{fig:sub3}
    \end{subfigure}

    \caption{Curve fitting models for the EfficientViT-DJSCC-AWGN case.}
    \label{fig:three_subfigs2}
\end{figure*}




By applying the same techniques as in \cite{nguyen2025semantic}, we introduce the unified loss model as in \eqref{eq_2dfit}, where $N_{\sf c}$ is the number of terms.
However, the MATLAB curve fitting tool does not support the multi-dimensional fitting with the structure given in  \eqref{eq_2dfit}. To find the parameters $\mu_0, \mu_{i,j},  i=1:6, j=1:N_{\sf c}$, we employ the gradient descent optimization method. The details of this approach are provided in Algorithm~\ref{alg1}, where lines 4-5 to determine the fitting values $\boldsymbol{\xi}$ and the error between the fitted results $\boldsymbol{\xi}$ and the measured data in Tables~\ref{tab:results_DJSCC} and ~\ref{tab:results_EfficientViT_DJSCC}, which is denoted as $\boldsymbol{Y}$. Lines $6-26$ are to compute the gradients of $\xi$ with respect to $\boldsymbol{\mu}_i, i=0:6$, while lines $27$ and $28$ are to update $\boldsymbol{\mu}_i, i=0:6$. In this paper, we set $\alpha_0 = \alpha_1 = \alpha_2 = 10^{-4}$, $\alpha_3 = \alpha_4 = 10^{-9}$, and $\alpha_5=10^{-10}$.
 The fitting model parameters corresponding to the results illustrated in Figs.~\ref{fig:three_subfigs} and \ref{fig:three_subfigs2} are given in Tables~\ref{tab:results_merged_row} and \ref{tab:results_efficientvit_awgn_merged} \ref{tab:results_merged_row}, respectively. 

\begin{figure}[t] 
    \centering
    \includegraphics[width=0.4\textwidth]{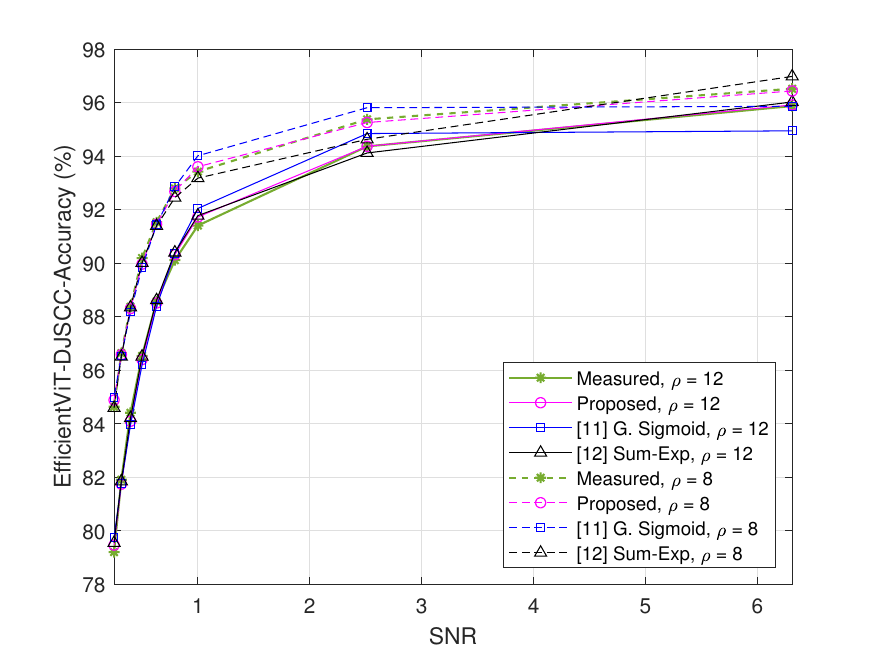}
    \captionsetup{font=small}
    \caption{A comparison of different fitting models.}
    \label{fig:comparison}
    \vspace{- 0.6 cm}
\end{figure}

Figs.~\ref{fig:three_subfigs} and \ref{fig:three_subfigs2} how that the proposed loss model fits well across all considered scenarios, including six different metrics (i.e., PSNR, MSE, SSIM, accuracy, precision, and recall) and two evaluation aspects (i.e., image reconstruction and image classification). Furthermore, we compare our approach with existing loss models to evaluate the fitting capability of the proposed unified loss model. In particular, the `G. Sigmoid' corresponds to the generalized sigmoid model in \cite{wang2025joint}, while the `Sum-Exp' represents the sum-of-exponential model in \cite{ding2023joint}. These models are defined as
\begin{equation}
\begin{aligned}
       \xi_{{\sf GSigmoid},\rho } &= \kappa_{1, \rho} + \frac{\kappa_{2, \rho}}{1 + \exp(\kappa_{3, \rho}s+\kappa_{4, \rho})}, \\
       \xi_{{\sf SumExp}, \rho} &= \kappa_{5,\rho} \exp(\kappa_{6, \rho}s) + \kappa_{7, \rho} \exp(\kappa_{8, \rho}s) + \kappa_{9, \rho},
\end{aligned}
\end{equation}
where $\kappa_{i, \rho}, \forall i$ are determined by fitting to empirical data corresponding to the compression ratio $\rho$. It means both $ \xi_{{\sf GSigmoid}, \rho }$ and $\xi_{{\sf SumExp}, \rho}$ are one-dimensional functions of the SNR $\gamma$, whereas our proposed model in \eqref{eq_2dfit} is a two-dimensional function that jointly depends on $\rho$ and $\gamma$. 
Since the existing models capture only one-dimensional behavior with respect to SNR $\gamma$, we evaluate them by fixing $\rho = 8$ and $\rho = 12$, and then optimally determining the corresponding parameters $\kappa_{i, \rho}$ for each case. There is no restriction on determining $\kappa_{i,8}$ and $\kappa_{i, 12}$. Fig.~\ref{fig:comparison} and Table~\ref{tab:compare} illustrate the accuracy results obtained by applying DJSCC with EfficientViT. Our proposed model significantly outperforms both the `G. Sigmoid' and `Sum-Exp' models, even while capturing both dimensions simultaneously.

\begin{table}[h]
\caption{Average MSE between the measured data and the fitted values.}
\label{tab:compare}
\centering
\begin{tabular}{|c|c|c|c|}
\hline
 & Proposed & G. Sigmoid & Sum-Exp. \\
\hline
$\rho=12$ & 0.0407 & 0.2474 & 0.0521 \\
\hline
$\rho=8$  & 0.0219 & 0.1425 & 0.1133 \\
\hline
\end{tabular}
\end{table}

\renewcommand{\arraystretch}{0.88}
\begin{table*}[!t]
\centering
\caption{\small\textsc{Parameters of the proposed model in Fig.~\ref{fig:three_subfigs} under DJSCC-AWGN conditions.}}
\label{tab:results_merged_row}
\begin{tabular}{@{}r>{\raggedleft\arraybackslash}p{1.2cm}*{6}{>{\raggedleft\arraybackslash}p{1.3cm}}@{}}
\toprule
\( i \backslash \text{Parameters} \) &
$\mu_{0}$ &
$\mu_{i,1}$ & $\mu_{i,2}$ & $\mu_{i,3}$ & $\mu_{i,4}$ &
$\mu_{i,5} \times 10^{3}$ & $\mu_{i,6}$ \\
\midrule

\multicolumn{8}{r}{\textbf{Image-MSE}} \\
\midrule
1 & \multirow{6}{*}{-132.851} &  325.366 & -308.162 & -2.900 & -2.386 &  -267.139 & 0.000 \\
2 &                       &  -206.275 & 549.240 & 2.640 & -0.632 &   61.433 & 0.000 \\
3 &                       &  -284.726 & -194.664 & -1.172 & -0.607 &  -328.019 & 0.000 \\
4 &                       &  -161.191 & 204.884 & -0.080 & 1.757 &   38.637 & 0.000 \\
5 &                       &   -65.899 & 61.225 & 2.840 & 5.172 &  222.567 & 0.000 \\
6 &                       &   929.199 & -1160.623 & 2.247 & 0.330 &   57.890 & 0.000 \\
\midrule

\multicolumn{8}{r}{\textbf{Image-PSNR}} \\
\midrule
1 & \multirow{6}{*}{29.574} &  36.649 & -3.168 & 0.632 & -3.436 &  -39.516 & 0.000 \\
2 &                       & -104.058 & -4.692 & 0.156 & -13.865 &  -12.833 & 0.000 \\
3 &                       &  -11.463 & 65.229 & 2.657 & 3.110 &   22.610 & 0.000 \\
4 &                       &  -55.011 & -4.879 & -0.433 & 1.825 &   19.184 & 0.000 \\
5 &                       &  -26.117 & 28.747 & -0.269 & 20.030 &   -7.904 & 0.000 \\
6 &                       &   75.898 & -14.652 & -2.346 & -2.705 &   -7.300 & 0.000 \\
\midrule

\multicolumn{8}{r}{\textbf{Image-SSIM}} \\
\midrule
1 & \multirow{6}{*}{28.768} &  126.256 & -15.514 & -1.740 & 19.184 & -308.684 & 0.000 \\
2 &                       &   55.943 & -11.283 & -0.996 & 0.955 &   19.585 & 0.000 \\
3 &                       &  153.574 & 4.197 & -0.213 & -7.228 & -701.928 & 0.000 \\
4 &                       & -206.346 & 16.021 & -1.375 & 1.974 & -548.854 & 0.000 \\
5 &                       &  -37.987 & 6.056 & 0.664 & 5.817 & -339.823 & 0.000 \\
6 &                       &  -33.014 & 28.336 & 4.731 & 0.568 &   78.149 & 0.000 \\
\bottomrule
\end{tabular}
\end{table*}

\begin{table*}[!t]
\centering
\caption{\small\textsc{Parameters of the proposed model in Fig.~\ref{fig:three_subfigs2} under EfficientViT-DJSCC-AWGN conditions.}}
\label{tab:results_efficientvit_awgn_merged}
\begin{tabular}{@{}r>{\raggedleft\arraybackslash}p{1.2cm}*{6}{>{\raggedleft\arraybackslash}p{1.3cm}}@{}}
\toprule
\( i \backslash \text{Parameters} \) &
$\mu_{0}$ &
$\mu_{i,1}$ & $\mu_{i,2}$ & $\mu_{i,3}$ & $\mu_{i,4}$ &
$\mu_{i,5} \times 10^{3}$ & $\mu_{i,6}$ \\
\midrule

\multicolumn{8}{r}{\textbf{EfficientViT-Accuracy}} \\
\midrule
1 & \multirow{6}{*}{62.683} & -128.798 &  -7.955 &  0.859 &  2.765 &  -1.475 &  0.008 \\
2 &                       &  -16.487 &  63.019 &  0.253 &  3.622 & -11.792 &  0.166 \\
3 &                       & -124.914 &   4.798 &  4.230 &  1.162 &  -7.734 & -0.069 \\
4 &                       &  146.847 &  20.979 & -1.989 & -0.866 & -14.133 & -0.112 \\
5 &                       &   20.938 & -106.143 & -3.172 & -2.294 &   5.252 &  0.113 \\
6 &                       &   63.187 &  15.759 &  0.622 &  4.974 &   2.075 & -0.000 \\
\midrule

\multicolumn{8}{r}{\textbf{EfficientViT-Precision}} \\
\midrule
1 & \multirow{6}{*}{62.695} & -128.788 &  -8.028 &  0.859 &  2.765 &  -1.316 &  0.008 \\
2 &                       &  -16.544 &  62.923 &  0.253 &  3.621 & -12.244 &  0.166 \\
3 &                       & -124.869 &   4.712 &  4.230 &  1.162 &  -8.155 & -0.069 \\
4 &                       &  146.910 &  21.145 & -1.989 & -0.866 & -13.233 & -0.112 \\
5 &                       &   20.899 & -106.007 & -3.172 & -2.294 &   4.992 &  0.112 \\
6 &                       &   63.198 &  15.761 &  0.622 &  4.974 &   2.005 & -0.000 \\
\midrule

\multicolumn{8}{r}{\textbf{EfficientViT-Recall}} \\
\midrule
1 & \multirow{6}{*}{62.667} & -128.816 &  -7.989 &  0.859 &  2.765 &  -0.928 &  0.008 \\
2 &                       &  -16.551 &  62.961 &  0.253 &  3.622 & -12.896 &  0.166 \\
3 &                       & -124.907 &   4.735 &  4.230 &  1.162 &  -8.707 & -0.069 \\
4 &                       &  146.867 &  21.114 & -1.989 & -0.866 & -11.932 & -0.113 \\
5 &                       &   20.886 & -106.115 & -3.172 & -2.294 &   4.889 &  0.112 \\
6 &                       &   63.170 &  15.741 &  0.622 &  4.974 &   1.814 & -0.000 \\
\bottomrule
\end{tabular}
\end{table*}
\renewcommand{\arraystretch}{1}

\section{Conclusions}
This work presents a novel data-fitting framework for modeling SC with DJSCC for EO imagery. In particular, it captures the relationship among the EO objective, compression ratio, and channel transmission conditions. By integrating real-world datasets and application-specific insights, the proposed framework provides a comprehensive approach for modeling SC with DJSCC in EO scenarios. Furthermore, comparisons with existing loss models confirm the strong fitting capability and superior performance of the proposed unified model.


\bibliographystyle{IEEEtran}
\bibliography{refs}
\end{document}